\documentclass[10pt,twocolumn,letterpaper]{article}

\usepackage[pagenumbers]{cvpr} 

\usepackage{booktabs}
\usepackage{xcolor}
\usepackage{colortbl}
\definecolor{hlRow}{HTML}{F3F6F9} 

\renewcommand{\paragraph}[1]{\vspace{.5em}\noindent\textbf{#1}}

\usepackage{algorithm}
\usepackage{algorithmic}
\usepackage{array}
\usepackage{multirow}
\usepackage{placeins}
\newcommand{\method}{\textsc{ExRel}}

\definecolor{cvprblue}{rgb}{0.21,0.49,0.74}
\usepackage[pagebackref,breaklinks,colorlinks,allcolors=cvprblue]{hyperref}

\def\confName{CVPR}
\def\confYear{2027}

\title{\method{}: Learning Directed Reuse Relevance between Rich-Text K--12 Question Images with Executable Relation Contracts}

\newcommand\blfootnote[1]{%
  \begingroup
  \renewcommand\thefootnote{}\footnote{#1}%
  \addtocounter{footnote}{-1}%
  \endgroup
}

\author{
Xuexin Liu$^{1*}$\quad
Zengyi Wo$^{2*\dagger}$\quad
Shiyu Zhang$^{2}$\quad
Suqi Cheng$^{3}$\quad
Daiting Shi$^{3}$\quad
Xiuyi Chen$^{3\dagger}$\\[0.4em]
$^{1}$Institute of Automation, Chinese Academy of Sciences\quad
$^{2}$Nanjing University\quad
$^{3}$Baidu
}

\begin{document}
\maketitle
\blfootnote{$^{*}$Equal contribution.\quad $^{\dagger}$Corresponding authors.}
\begin{abstract}
In K--12 education, question images combine text, mathematical expressions, diagrams, and diverse layouts. Variations in wording, conditions, answer options, or task scope can substantially alter their reuse utility despite visual similarity. We formulate this problem as \emph{directed reuse relevance}, grading candidate utility as Strong, Medium, or Weak. However, generic similarity scores overlook task-specific reuse requirements, while natural supervision is dominated by Strong relations and lacks fine-grained coverage. We propose \method{}, an agentic, code-grounded framework for controllable and verifiable visual data synthesis. It reconstructs real questions into editable HTML/CSS seeds and compiles 15 relation subtypes into executable contracts specifying permitted edits, protected invariants, and query--candidate direction. Synthesized pairs undergo deterministic DOM/render validation and target-hidden, evidence-assisted review, with bounded repair before mild photographic degradation. Only rendered images are used for downstream training. Experiments show that \method{} supervision outperforms size-matched natural data and count- and class-matched replay, particularly on Medium relations, while maintaining comparable query-level ranking performance and improving robustness to visual degradation. Construction ablations demonstrate higher valid yield and fewer false admissions than less structured baselines. Human audits confirm 98.0\% post-admission validity, while \method{} achieves a lower per-accepted-pair cost than outsourced annotation (0.70 vs.\ 1.10 RMB). Downstream answer routing further demonstrates the practical utility of the synthesized supervision.
\end{abstract}

\section{Introduction}
\label{sec:intro}

Retrieving existing K--12 question images to support exercise design, homework assembly, and solution reference is a routine workflow in educational platforms. Unlike general document retrieval~\cite{faysse2025colpali,dong2025mmdocir}, where topical or visual similarity often suffices, educational reuse hinges on fine-grained task compatibility: two visually similar question images may share the same diagram and layout yet differ in a single numerical condition, option set, or subquestion scope, completely changing whether the retrieved candidate can be reused directly, adapted with edits, or discarded. We formulate this decision as \emph{directed reuse relevance}, assigning each ordered query--candidate image pair a grade of Strong, Medium, or Weak.

\begin{figure*}[t]
\centering
\includegraphics[width=0.99\textwidth,height=0.31\textheight,keepaspectratio]{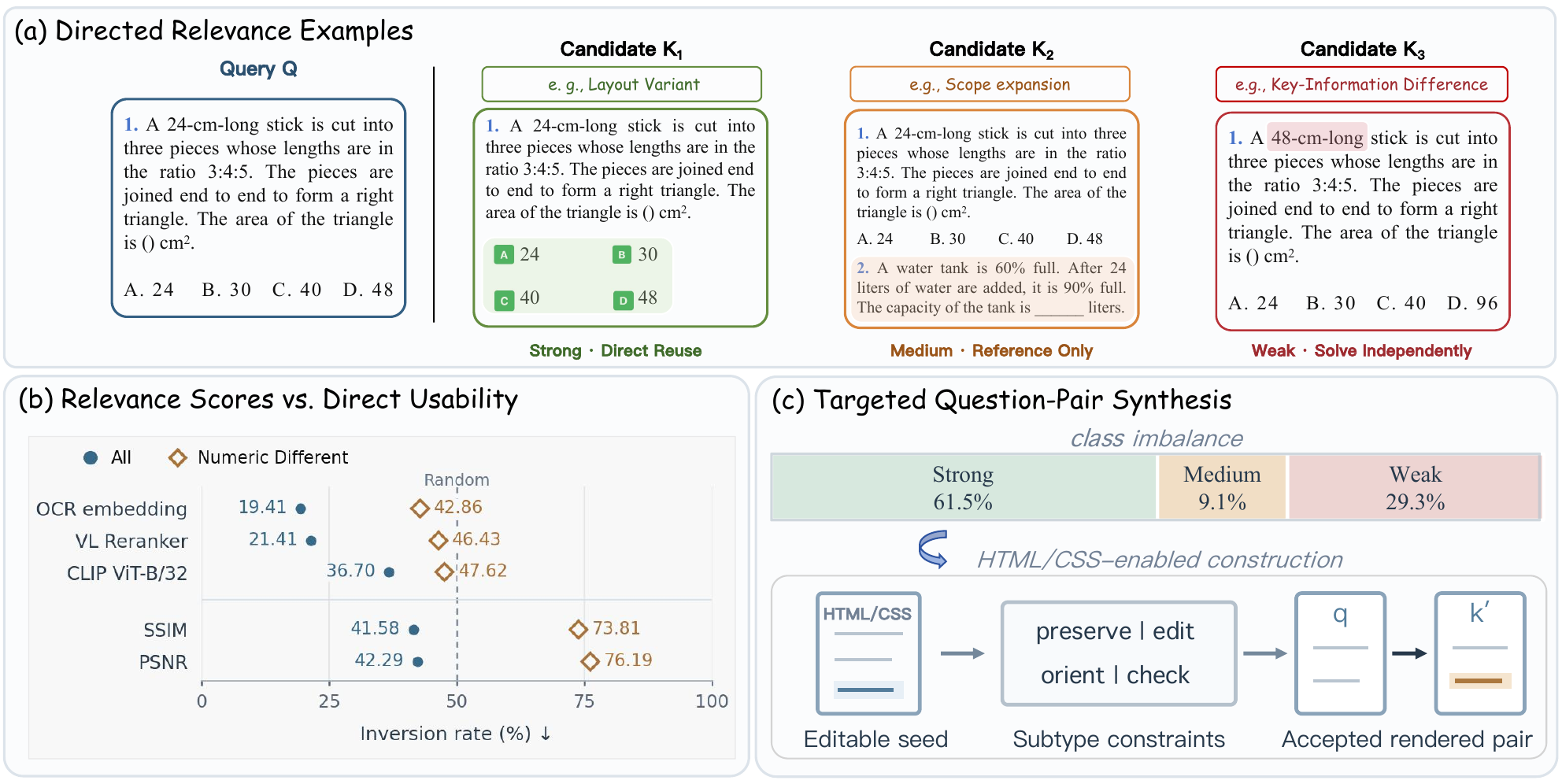}
\caption{\textbf{Motivation and overview of directed reuse relevance.}
(a)~Examples of Strong, Medium, and Weak relations.
(b)~Strong--Weak ranking inversion rates of similarity-based metrics and multimodal rankers.
(c)~Class imbalance in natural supervision motivates contract-guided synthesis from editable HTML/CSS seeds.}
\label{fig:overview}
\end{figure*}

Reuse depends on \emph{what} changes and \emph{in which direction}, rather than visual distance alone (Figure~\ref{fig:overview}(a)). A layout change may preserve direct reuse (Strong), whereas changes to critical conditions or answer options may require adaptation (Medium) or invalidate reuse (Weak). Relevance is also directional: a candidate containing the complete query plus an additional compatible subquestion may be Medium, while reversing the pair can make it Weak. Our preliminary analysis further reveals substantial Strong--Weak ranking inversions by pixel-based metrics and multimodal rankers, particularly for numeric-difference pairs (Figure~\ref{fig:overview}(b)). These observations motivate supervision that captures fine-grained reuse boundaries.

Beyond this modeling challenge, naturally collected pairs provide uneven coverage: Strong pairs dominate, leaving Medium and Weak relations underrepresented (Figure~\ref{fig:overview}(c)). Although replaying minority pairs can rebalance class frequencies, it introduces limited diversity beyond existing realizations. Improving subtype coverage therefore requires targeted synthesis of relation-specific question pairs.

Such synthesis poses two challenges. First, a target grade alone does not specify a valid edit: unconstrained generation may modify unintended content, remove critical conditions, introduce rendering defects, or reverse query--candidate direction. Second, existing generation and verification approaches do not directly address the joint requirements of fine-grained semantic editing, structural preservation, and directional relation verification. In particular, editable web representations~\cite{laurencon2024websight,si2025design2code,li2026puredocbench} expose structural elements but do not by themselves ensure that edits preserve the intended reuse relation. These challenges call for a framework that jointly controls semantic interventions, preserves document structure, and verifies the resulting relations.

To address these challenges, we propose \method{}, an \emph{agentic, code-grounded framework for controllable and verifiable visual data synthesis}. Rather than generating question images from scratch, \method{} converts real question images into editable HTML/CSS seeds, profiles their structural slots, and compiles each target subtype into an executable \emph{relation contract} that governs what may be edited, what must remain invariant, and how the query--candidate pair is oriented. An agentic controller then executes the intervention and gates admission by combining deterministic DOM/render checks with target-hidden review, applying bounded repair before mild photographic degradation. All structural scaffolding is discarded after admission so that the downstream predictor learns purely from rendered image pairs.

We evaluate \method{} on the human-adjudicated Release-200 benchmark (1,000 real pairs across 200 queries) and the subtype-balanced Uniform-300 diagnostic set. Across both fixed and additive training budgets, \method{} supervision consistently improves real-image relevance classification---most notably on the scarce Medium adaptation boundary---while preserving query-level ranking quality and strengthening resilience to capture degradation. Ablation, human-audit, cost, and answer-routing studies further verify that contract-guided construction yields higher-fidelity supervision at lower unit cost than outsourced annotation and translates into effective downstream question answering. Our contributions are:

\begin{itemize}[leftmargin=*,nosep]

\item \textbf{Directed reuse relevance and benchmark.}
We formulate directed reuse relevance as a three-grade, asymmetric prediction task with 15 fine-grained subtypes, and establish Release-200, a human-adjudicated benchmark of real question-image pairs.

\item \textbf{Agentic, code-grounded data synthesis.}
We propose \method{}, a controllable and verifiable visual data synthesis framework that combines agentic planning, editable HTML/CSS, and executable relation contracts to generate reliable, fine-grained supervision.

\item \textbf{Data quality, efficiency, and downstream utility.}
We demonstrate the effectiveness and efficiency of synthesized supervision through predictive evaluations, construction analyses, human audits, cost comparisons, and downstream answer routing.

\end{itemize}

\section{Related Work}
\label{sec:related}

\subsection{Relevance Modeling for Multimodal Retrieval}

Question retrieval ranks existing questions by semantic relevance~\cite{cai2021questionretrieval}, while duplicate detection identifies questions expressing the same information need~\cite{ruckle2019duplicate}. Graded retrieval methods such as GR$^2$~\cite{tang2024gr2} and MERGE~\cite{zhang2025merge} incorporate multi-level relevance into training, while EXAM~\cite{farzi2024answerability} evaluates retrieved content by its utility for answering exam questions. For visually rich documents, ColPali~\cite{faysse2025colpali} embeds page images directly, MMDocIR~\cite{dong2025mmdocir} benchmarks page- and layout-level retrieval, and MM-R5~\cite{xu2025mmr5} introduces reasoning into multimodal reranking. However, these methods primarily address semantic matching or document retrieval, rather than the \emph{directed reuse utility} between visually similar question images. In K--12 question reuse, minor changes in mathematical tokens, numerical values, or question scope can alter whether a candidate supports direct reuse, requires adaptation, or should be rejected. Moreover, reversing the query--candidate order may change the relevance grade. \method{} addresses these challenges by modeling graded, directed reuse relevance with supervision tailored to question-image pairs.

\subsection{Agentic and Verified Synthetic Data Generation}

Recent synthetic data pipelines combine model generation with filtering, feedback, and verification. Self-Instruct~\cite{wang2023selfinstruct} filters generated instructions, while Promptagator~\cite{dai2023promptagator} uses round-trip consistency for query generation. AgentInstruct~\cite{mitra2024agentinstruct} and Autodata~\cite{kulikov2026autodata} employ agentic workflows, while DataEnvGym~\cite{khan2025dataenvgym} explores feedback-driven generation. For data verification, Benchmark Agent~\cite{xiong2026benchmarkagent} combines tool execution with quality checks, DynaMath~\cite{zou2025dynamath} generates visual math variants from programs, and TrustGeoGen~\cite{fu2025trustgeogen} applies formal verification to geometry problems. Meanwhile, WebSight~\cite{laurencon2024websight}, Design2Code~\cite{si2025design2code}, and PureDocBench~\cite{li2026puredocbench} explore editable web representations for structural control over visual content. Despite these advances, synthesizing \emph{graded, directed relations} between visual question instances under fine-grained semantic and structural constraints remains underexplored. \method{} addresses this gap by using executable relation contracts to coordinate structure-aware HTML/CSS editing, semantic interventions, and evidence-based admission.

\section{Method}
\label{sec:method}


\subsection{Problem Formulation}
\label{sec:formulation}

\paragraph{Directed reuse relevance.}
Given a query image $q$ and a retrieved candidate image $k$, \emph{directed reuse relevance} $y(q\!\rightarrow\!k)\in\mathcal{Y}=\{\mathrm{Strong},\mathrm{Medium},\mathrm{Weak}\}$ measures how well $k$ supports the task posed by $q$. Strong indicates direct reusability; Medium denotes useful but adaptation-dependent support; and Weak indicates insufficient support or a critical mismatch. These grades capture task-reuse utility rather than visual similarity or answer correctness. For example, a semantically equivalent rephrase or option-order change is Medium because reuse requires adapting the presentation or remapping option-indexed answers.

Relevance is \emph{directional}. If $k$ contains the complete question in $q$ plus an additional compatible subquestion (\emph{extra-question containment}), it is Medium because the query can be supported by selecting the relevant content from $k$. Reversing the pair yields \emph{partial query containment} and is Weak because the candidate no longer covers the full task in the query. Thus, $y(q\!\rightarrow\!k)$ need not equal $y(k\!\rightarrow\!q)$. The downstream predictor $f_\theta$ takes the ordered rendered pair $(q,k)$ and outputs $\hat{y}=f_\theta(q,k)\in\mathcal{Y}$.

\begin{table}[t]
\centering
\small
\setlength{\tabcolsep}{7pt}
\renewcommand{\arraystretch}{1.08}
\caption{\textbf{Two-level directed reuse taxonomy.} Three operational classes serve as prediction targets; 15 subtypes specify finer-grained relation patterns and construction constraints.}
\label{tab:taxonomy}
\begin{tabular}{l|l}
\toprule
\textbf{Class} & \textbf{Subtypes} \\
\midrule
\multirow{3}{*}{Strong}
& Exact-same layout variant \\
& Noncritical extra context \\
& Complete same-question containment \\
\midrule
\multirow{6}{*}{Medium}
& Extra-question containment \\
& Material/cloze completion \\
& Semantic-equivalent rephrase \\
& Question-type variant \\
& Option-order variant \\
& Partial-option content variant \\
\midrule
\multirow{6}{*}{Weak}
& Numeric difference \\
& Key-information difference \\
& Partial query containment \\
& Similar/different question \\
& Same-material/different question \\
& Hard-negative retrieval pair \\
\bottomrule
\end{tabular}
\end{table}

\paragraph{Taxonomy and synthesis objective.}
The taxonomy in Table~\ref{tab:taxonomy} decomposes the three operational classes into 15 finer-grained relation subtypes and assigns each subtype $s\in\mathcal{S}$ to one target class. Let $\mathcal{S}_c$ denote the subtypes assigned to class $c\in\mathcal{Y}$; these sets form a partition $\mathcal{S}=\bigcup_{c\in\mathcal{Y}}\mathcal{S}_c$ with $\mathcal{S}_c\cap\mathcal{S}_{c'}=\emptyset$ ($c\neq c'$), and $y_s=c$ if $s\in\mathcal{S}_c$. Given a construction target $\tau$, an executable seed $x$ (Section~\ref{sec:seed}), and an eligible subtype $s$, \method{} performs conditional synthesis:
\begin{equation}
(p,z)=\mathcal{G}(x,s;\mathcal{B}),\qquad
p\in\bigl\{(q,k,y_s),\,\bot\bigr\},
\label{eq:generator}
\end{equation}
where $\mathcal{B}$ is a source-question pool used by subtypes requiring a second source, $\bot$ denotes rejection, and $z$ records the instantiated contract, edit evidence, validation outcomes, and terminal state. The objective is to produce diverse realizations that satisfy the semantic, structural, directional, and visual constraints of subtype $s$, rather than to generate unconstrained questions.

\subsection{Overview of \method{}}
\label{sec:overview}

\begin{figure*}[t]
\centering
\includegraphics[width=0.99\textwidth,height=0.4\textheight,keepaspectratio]{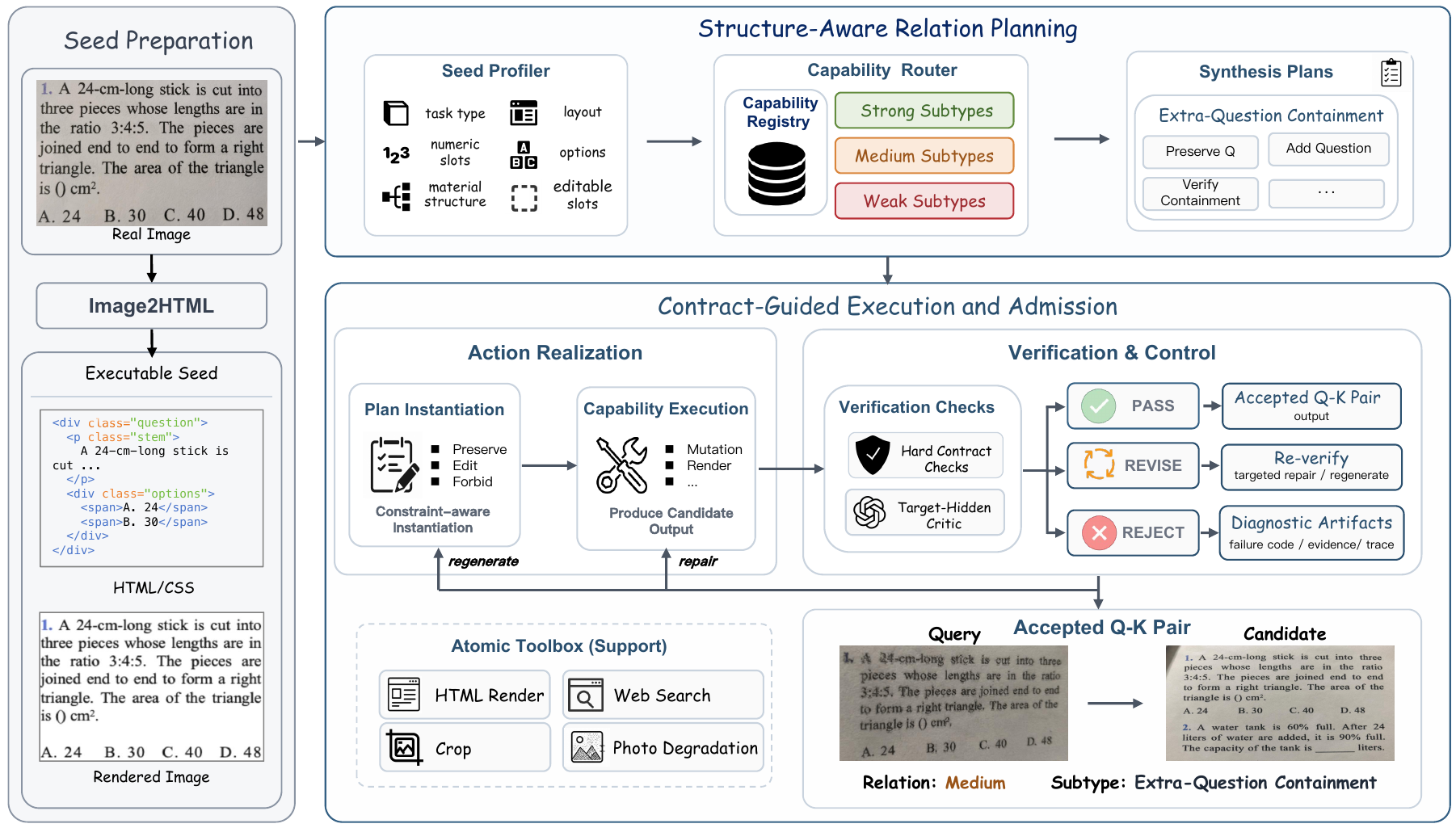}
\caption{\textbf{Contract-guided synthesis in \method{}.} A real question image is reconstructed as an editable seed. The planner profiles its structure, selects an eligible subtype matching the construction target, and instantiates its executable contract. The executor applies the specified intervention and combines deterministic validation with target-hidden review; failed candidates undergo bounded revision or rejection.}
\label{fig:method}
\end{figure*}

As illustrated in Figure~\ref{fig:method}, the framework comprises three stages: (1)~\emph{Executable Seed Preparation} reconstructs each image as editable HTML/CSS and verifies its canonical render (Section~\ref{sec:seed}); (2)~\emph{Structure-Aware Relation Planning} identifies feasible subtypes and instantiates their contracts (Section~\ref{sec:planning}); and (3)~\emph{Contract-Guided Execution and Admission} performs the planned intervention and verifies the resulting pair (Section~\ref{sec:execution}). Agents interpret target relations and coordinate the workflow, while code-level operations precisely manipulate editable document elements. The complete synthesis procedure is summarized in Algorithm~\ref{alg:exrel} (Appendix~\ref{sec:app_taxonomy}).

\subsection{Executable Seed Preparation}
\label{sec:seed}

\method{} starts from real question images rather than generating visual content from scratch. Given a real image $I^{\mathrm{real}}$, Image2HTML reconstructs an editable HTML/CSS representation $H$, which is rendered in a browser to obtain its canonical view $I=\mathrm{Render}(H)$. This conversion enables subsequent interventions at the document-structure level rather than through direct pixel manipulation.

The reconstruction agent checks the rendered question body for missing text, clipping, overlap, and unreadable regions. Handwriting, user annotations, and capture noise are excluded from this reconstruction check. Failed reconstructions are dropped or retried within a bounded budget. The resulting executable seed is $x=(I,H,m)$, where $m$ stores structural and provenance metadata, including source identity for split control. The DOM representation exposes slots for stems, options, and layout elements, providing the structural basis for subtype feasibility analysis and edit verification.

\subsection{Structure-Aware Relation Planning}
\label{sec:planning}

\paragraph{Seed profiling and subtype routing.}
The profiler extracts question and material boundaries, options, blanks, numeric tokens, task anchors, geometry, editable DOM slots, source identity, and subtype eligibility into metadata $m$. For example, an option-order variant requires separable option nodes, whereas a numeric-difference variant requires identifiable numerical tokens in the stem. Each subtype precondition $P_s(x;\mathcal{B})\in\{0,1\}$ determines whether seed $x$ supports subtype $s$; for two-source subtypes such as hard negatives, the check also considers the source pool $\mathcal{B}$. The eligible subtype set is
\begin{equation}
\mathcal{E}(x;\mathcal{B})=\bigl\{s\in\mathcal{S}\mid P_s(x;\mathcal{B})=1\bigr\}.
\label{eq:eligible}
\end{equation}
The planner selects $s\in\mathcal{E}(x;\mathcal{B})$ according to construction target $\tau$. If no eligible subtype matches $\tau$, the seed is rejected.

\paragraph{Executable relation contracts.}
For each subtype $s$, \method{} instantiates an executable contract
\begin{equation}
\mathcal{C}_s=(P_s,\,A_s,\,\Phi_s,\,D_s,\,V_s,\,R_s),
\label{eq:contract}
\end{equation}
where $P_s$ specifies structural preconditions, $A_s$ defines permitted interventions, $\Phi_s$ specifies protected semantic and structural invariants, $D_s$ determines query--candidate direction, $V_s$ defines admission checks, and $R_s$ specifies the repair policy. Instantiating $\mathcal{C}_s$ on seed $x$ binds editable and protected DOM regions into an execution plan $\pi$ that translates the target relation into concrete editing and verification constraints.

For \emph{extra-question containment}, for example, $P_s$ requires a complete question and an appendable region; $A_s$ permits adding a compatible subquestion; $\Phi_s$ preserves the original stem; and $D_s$ assigns the original question to $q$ and the extended version to $k$. The admission checks $V_s$ assess containment, layout integrity, and consistency with the Medium relation, while $R_s$ bounds repair or regeneration.

\subsection{Contract-Guided Execution and Admission}
\label{sec:execution}

Given execution plan $\pi$, \method{} uses code-level operators to realize the specified intervention and applies complementary checks to determine whether the resulting pair is admissible. This stage combines precise substrate-level control with evidence-assisted assessment of the rendered relation.

\paragraph{Intervention execution and pair assembly.}
The executor applies $\pi$ to the specified DOM slots. Deterministic operators perform programmatic edits, such as option permutation, bounded cropping, and literal replacement, while VLM operators handle semantic interventions, such as modifying a key condition, target quantity, or subquestion. The executor returns modified HTML $H'$, edit record $e$, and rendered image $I'=\mathrm{Render}(H')$. For hard-negative pairs, the controller selects two distinct, unedited source records from $\mathcal{B}$ and records their identities in $e$ for distinctness verification. The direction specified by $D_s$ determines the ordered pair $(q,k)$.

\paragraph{Evidence-aligned admission.}
Admission requires both compliance with the specified intervention and consistency of the resulting relation. Deterministic validators $V_s$ produce Boolean checks
$\mathbf{v}=(v_s^{\mathrm{dom}},v_s^{\mathrm{ren}},v_s^{\mathrm{qual}})\in\{0,1\}^3$
for contract compliance, render integrity, and visual quality. In parallel, a target-hidden, evidence-assisted reviewer predicts $\tilde{y}\in\mathcal{Y}$ using the ordered rendered image pair $(q,k)$ together with observable HTML evidence and the relation diff in edit record $e$. The reviewer is not explicitly provided with the target subtype $s$ or target label $y_s$. A pair is admitted iff
\begin{equation}
\mathrm{Accept}_s(\mathbf{v},\tilde{y})=
v_s^{\mathrm{dom}}\wedge v_s^{\mathrm{ren}}
\wedge v_s^{\mathrm{qual}}\wedge\mathbf{1}[\tilde{y}=y_s].
\label{eq:accept}
\end{equation}

The deterministic validators enforce process compliance: $v_s^{\mathrm{dom}}$ checks DOM snapshots and edit diffs against $A_s$ and $\Phi_s$, while $v_s^{\mathrm{ren}}$ and $v_s^{\mathrm{qual}}$ assess render integrity and visual readability, respectively. The reviewer uses the rendered pair and associated construction evidence to assess relation consistency. A reviewer agreement cannot override a failed deterministic check, and any mismatch with $y_s$ leads to rejection. Post-admission validity is further assessed through the human audit in Table~\ref{tab:supervision_fidelity}.

\paragraph{Bounded feedback and photographic degradation.}
Failed checks trigger subtype-specific corrective actions under $R_s$. The synthesis loop allows at most one semantic regeneration after the initial attempt and at most one content-preserving render repair per semantic attempt; all retry and repair decisions are recorded in trace $z$.

After a pair passes clean-view admission, a deterministic, seed-controlled pipeline generates a mildly degraded counterpart using sampled effects such as uneven lighting, geometric distortion, noise and blur, compression, low-quality printing, and ink bleed. The sampled scene and parameters are recorded for each pair. The degraded counterpart is retained as an additional training view and does not undergo a separate admission review.

\paragraph{Downstream relevance predictor training.}
The admitted clean and corresponding degraded views form the synthetic supervision set $\mathcal{D}_{\mathrm{syn}}$. We train the downstream predictor $f_\theta$ on $\mathcal{D}_{\mathrm{syn}}$, either alone or together with natural supervision $\mathcal{D}_{\mathrm{real}}$, using cross-entropy over the three grades in $\mathcal{Y}$. After admission, HTML/DOM structures, edit records, and contract traces are discarded. Thus, structured representations support data construction and verification, while both training and inference operate exclusively on ordered rendered image pairs $(q,k)$.
\section{Experiments}
\label{sec:experiments}

We evaluate \method{} on three questions: (i)~Does synthetic supervision improve directed relevance prediction on real question images (Section~\ref{sec:effectiveness})? (ii)~Does contract-guided construction raise valid yield and reduce false admissions over less structured baselines (Section~\ref{sec:analysis})? (iii)~Do predicted grades support downstream question answering (Section~\ref{sec:downstream})?

\subsection{Experimental Settings}
\label{sec:setup}

\begin{figure}[htb]
    \centering
    \includegraphics[width=0.48\textwidth]{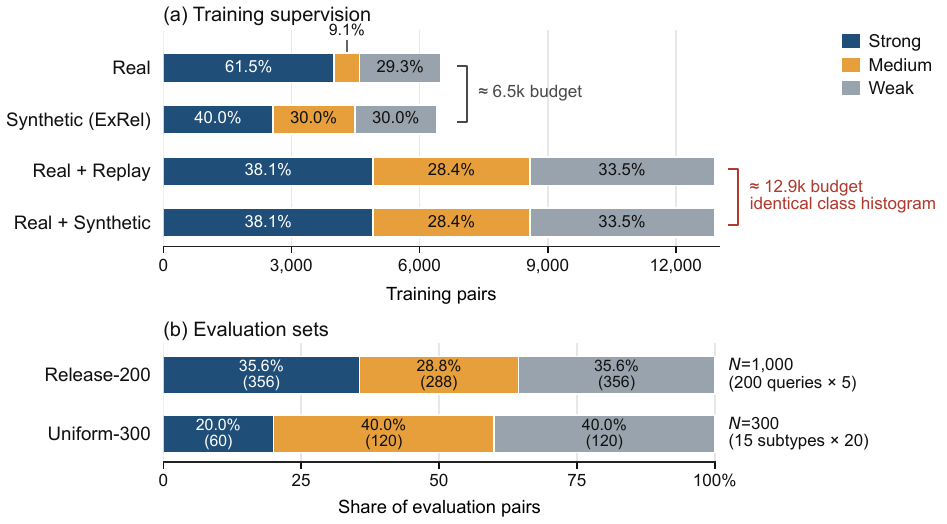}
    \caption{\textbf{Class distributions} of training supervision and evaluation benchmarks. Additive settings match in sample count and class proportions.}
    \label{fig:data_distribution}
\end{figure}

\paragraph{Datasets.}
We study two sources of training supervision: $\approx$6.5K naturally collected query--candidate pairs (\textit{Real}) and 6.4K pairs synthesized by \method{} (\textit{Synthetic}). Both follow the three-grade taxonomy, while \method{} explicitly targets all 15 subtypes. As shown in Figure~\ref{fig:data_distribution}, natural pairs are heavily skewed toward Strong relations (61.5\%), whereas synthetic pairs provide balanced Medium and Weak coverage. For evaluation, we use two benchmarks: \textbf{Release-200} comprises 200 real-image queries paired with five retrieved candidates each (1,000 ordered pairs) with human-adjudicated labels; \textbf{Uniform-300} contains 20 pairs per subtype (300 total), each evaluated in a clean browser-rendered view and a paired mildly degraded view sharing the same pair IDs to diagnose in-pipeline subtype coverage and degradation robustness. To prevent data leakage, we exclude all 350 held-out evaluation queries (comprising Release-200 and 150 reserve queries) and their associated source questions before constructing the synthesis seed pool, yielding 6,700 validated reconstructed seeds.

\paragraph{Baselines.}
We compare \method{} against three groups of baselines:
(1)~\emph{Zero-shot VLMs}, including proprietary models (GPT-5.5~\cite{openai2026gpt55}, Claude Opus 4.7~\cite{anthropic2026opus47}, Claude Sonnet 4.6~\cite{anthropic2026sonnet46}, Doubao Seed 2.1 Pro/2.0 Lite~\cite{volcengine2026models}) and open-weight models (Qwen3.5-4B/27B~\cite{qwen2026qwen35}, where 4B is our untuned backbone).
(2)~\emph{Supervised data baselines} on Qwen3.5-4B: a \emph{fixed-budget} setting comparing \textit{Synthetic (\method{})} against size-matched \textit{Real}, and an \emph{additive} setting ($\approx$12.9K pairs) comparing \textit{Real\,+\,Synthetic (\method{})} against \textit{Real\,+\,Replay} (which augments \textit{Real} with real pairs re-sampled to match the synthetic count and class distribution, isolating new subtype realizations from class rebalancing).
(3)~\emph{Construction baselines} (Section~\ref{sec:analysis}), comparing \method{} against \emph{Image-based} editing (direct pixel manipulation) and \emph{HTML-enabled} editing (reconstructed HTML/CSS without contracts).

\paragraph{Evaluation protocol and metrics.}
Every model predicts a grade $\hat{y}\in\mathcal{Y}$ from an ordered pair $(q,k)$. On Release-200, we report pair-level accuracy (Acc.), Macro-F1 (mF1), and per-class F1 for Strong, Medium, and Weak (S-F1, M-F1, W-F1). At the query level, predicted grades rank each query's candidates as Strong $\succ$ Medium $\succ$ Weak; we report tie-aware nDCG@5~\cite{jarvelin2002cumulated} (gains $2,1,0$) and \emph{Exact-5} ($\frac{1}{|\mathcal{Q}|}\sum_{q\in\mathcal{Q}}\prod_{j=1}^{5}\mathbf{1}[\hat{y}_{q,j}=y_{q,j}]$, the fraction of queries with all five candidates graded correctly). On Uniform-300, the clean and degraded views are evaluated separately for pair-level accuracy. Comparisons with additional OCR-based and vision--language rankers~\cite{zhang2025qwen3embedding,chen2024bge,baai2024bgereranker,li2026qwen3vlembedding} using Strong-based Hit@1, MRR@5, and R@3 are reported in Appendix~\ref{sec:app_ranking} (Table~\ref{tab:off_the_shelf_ranking}). The Strong--Weak inversion rate in Figure~\ref{fig:overview}(b) (including CLIP ViT-B/32~\cite{radford2021clip}, SSIM~\cite{wang2004ssim}, and PSNR) scores 1 for an inversion, 0.5 for a tie, and 0 otherwise, averaged within and across eligible queries.

\paragraph{Implementation details.}
In our synthesis pipeline, Image2HTML reconstruction, structure-aware planning, semantic editing, repair calls, and target-hidden paired review use GPT-5.5~\cite{openai2026gpt55} (with temperature zero for reconstruction, structured planning, and review). For downstream relevance prediction, all supervised models fine-tune Qwen3.5-4B~\cite{qwen2026qwen35} via \texttt{ms-swift}~\cite{zhao2024swift} with identical hyperparameters: full-parameter \texttt{bfloat16} optimization on NVIDIA A800 GPUs with DeepSpeed ZeRO-3~\cite{rajbhandari2019zero}, sequence length 4096, effective batch size 128, and learning rate $10^{-5}$ with cosine decay and 5\% linear warm-up.

\subsection{Main Results on Relevance Prediction}
\label{sec:effectiveness}


\begin{table*}[htb]
\centering
\caption{
    \textbf{Main results (\%) on Release-200 and Uniform-300.}
    We compare zero-shot baselines and supervised models under
    fixed and additive data budgets.
    Release-200 reports accuracy (Acc.), macro-F1 (mF1), per-class F1 for Strong, Medium, and Weak (S-F1, M-F1, W-F1),
    nDCG@5, and Exact-5; Uniform-300 reports accuracy under
    clean and degraded conditions.
    Higher is better for all metrics.
    \textbf{Bold} and \underline{underlined} indicate the
    best and second-best results, respectively.
}
\label{tab:main_results}

\fontsize{8}{9.5}\selectfont
\setlength{\tabcolsep}{3pt}
\renewcommand{\arraystretch}{1.08}

\begin{tabular*}{\textwidth}{
    @{\extracolsep{\fill}} l ccccccc c c @{}
}
\toprule
& \multicolumn{7}{c}{\textbf{Release-200}}
& \multicolumn{2}{c}{\textbf{Uniform-300}} \\
\cmidrule(lr){2-8}
\cmidrule(lr){9-10}

\textbf{Method}
& \textbf{Acc.}
& \textbf{mF1}
& \textbf{S-F1}
& \textbf{M-F1}
& \textbf{W-F1}
& \textbf{nDCG@5}
& \textbf{Exact-5}
& \textbf{Clean}
& \textbf{Degr.} \\
\midrule

\multicolumn{10}{@{}l}{
    \textbf{{Proprietary Zero-Shot Models}}
} \\
\addlinespace[1pt]

GPT-5.5
& 67.70 & 66.74 & 68.36 & 55.40 & 76.46
& 92.75 & 28.00 & 75.00 & 65.67 \\

Claude Opus 4.7
& 67.80 & 67.62 & 69.85 & 57.28 & 75.74
& 92.78 & 28.50 & --- & --- \\

Claude Sonnet 4.6
& 60.40 & 60.14 & 67.93 & 50.71 & 61.78
& 89.51 & 17.50 & 68.67 & 66.00 \\

Doubao Seed 2.1 Pro
& 63.90 & 59.29 & 73.05 & 34.17 & 70.66
& 92.74 & 30.00 & 71.00 & 68.00 \\

Doubao Seed 2.0 Lite
& 60.40 & 61.36 & 72.19 & 44.84 & 67.06
& 92.86 & 22.50 & 62.67 & 59.67 \\

\midrule

\multicolumn{10}{@{}l}{
    \textbf{{Open-Weight Zero-Shot Models}}
} \\
\addlinespace[1pt]

Qwen3.5-4B
& 63.40 & 61.07 & 70.68 & 40.17 & 72.36
& 92.85 & 20.00 & 66.67 & 63.67 \\

Qwen3.5-27B
& 69.60 & 69.18 & 75.11 & 57.29 & 75.14
& \underline{93.47} & 30.50 & 80.33 & 78.67 \\

\midrule

\multicolumn{10}{@{}l}{
    \textbf{{Fixed Budget Setting}}
} \\
\addlinespace[1pt]

Real
& 76.00 & 74.17 & 79.30 & 62.53 & 80.69
& 93.13 & 43.00 & 77.67 & 74.33 \\

\rowcolor{hlRow}
\textbf{Synthetic (\method{})}
& 76.90 & 75.58 & 80.15 & 65.98 & 80.63
& 93.27 & 43.00
& \underline{90.67} & \underline{90.67} \\

\midrule

\multicolumn{10}{@{}l}{
    \textbf{{Additive Budget Setting}}
} \\
\addlinespace[1pt]

Real\,+\,Replay
& \underline{78.80}
& \underline{77.24}
& \underline{82.53}
& \underline{67.40}
& \underline{81.79}
& \underline{93.47}
& \textbf{47.00}
& 79.00
& 75.33 \\

\rowcolor{hlRow}
\textbf{Real\,+\,Synthetic (\method{})}
& \textbf{79.90}
& \textbf{78.72}
& \textbf{83.95}
& \textbf{70.20}
& \textbf{82.00}
& \textbf{93.52}
& \underline{45.50}
& \textbf{98.00}
& \textbf{96.00} \\

\bottomrule
\end{tabular*}
\end{table*}

\begin{figure}[htb]
    \centering
    \includegraphics[width=0.48\textwidth]{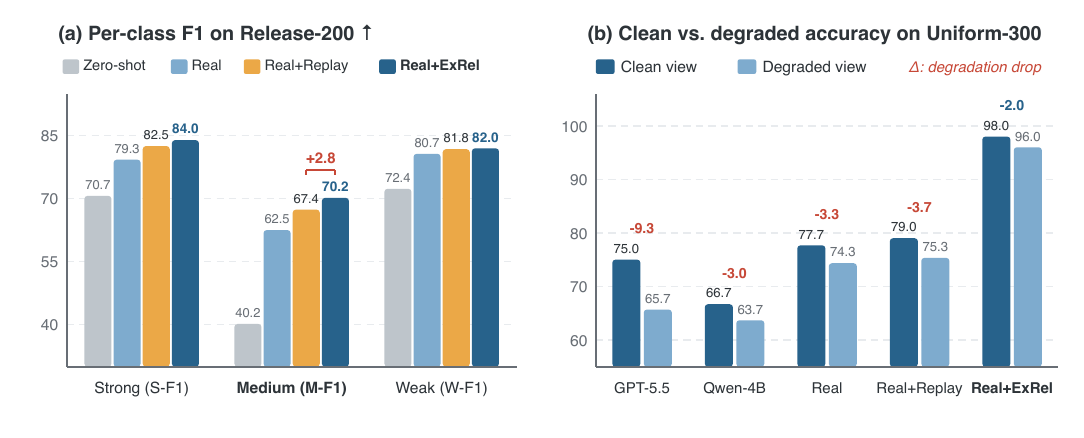}
    \caption{\textbf{Class-wise and degradation diagnostics.} (a)~Per-class F1 across Qwen3.5-4B training regimes on Release-200, highlighting the Medium adaptation boundary gain. (b)~Clean vs.\ degraded accuracy and degradation drop ($\Delta$) on Uniform-300.}
    \label{fig:regime_diagnostics}
\end{figure}

As shown in Table~\ref{tab:main_results}, we evaluate relevance prediction performance across zero-shot and supervised settings, from which we draw the following conclusions:
\begin{itemize}[leftmargin=*,nosep]
\item \textit{Zero-shot VLMs struggle on the Medium adaptation boundary.} Across all proprietary and open-weight zero-shot baselines, M-F1 is substantially lower than S-F1 and W-F1. Off-the-shelf VLMs rely on global visual and topical similarity: they frequently over-predict Strong when a single modified number invalidates direct reuse, and lack an operational boundary for when a candidate requires adaptation. Fine-tuning Qwen3.5-4B on our supervision yields a large leap over both its untuned backbone and the much larger Qwen3.5-27B baseline across pair-level metrics and Exact-5, underscoring the necessity of task-specific supervision.

\item \textit{Under the fixed budget, synthetic supervision alone outperforms natural supervision on real images.} On Release-200, training exclusively on \textit{Synthetic (\method{})} surpasses size-matched \textit{Real} in overall accuracy and mF1 (+1.41 mF1), with a marked advantage on the Medium class (+3.45 M-F1) while remaining on par on W-F1 and Exact-5. Because natural data under-expose the predictor to fine-grained content variations, systematic coverage of all 15 subtypes combined with photographic degradation more than compensates for any render-to-capture domain gap.

\item \textit{Under the additive budget, \method{} improves pair-level discrimination beyond class rebalancing.} Augmenting natural data with \method{} consistently outperforms \textit{Real\,+\,Replay} across accuracy, mF1 (+1.48), and all three per-class F1 scores, with the largest gain on the Medium boundary (+2.80 M-F1; Figure~\ref{fig:regime_diagnostics}(a)). Unlike replay, which repeats existing minority examples, \method{} introduces new structural and semantic boundary realizations around real seeds. At the query level, nDCG@5 is virtually tied as five-candidate ranking with coarse gains is near saturation, whereas Exact-5 slightly favors replay because heavier reinforcement of majority Strong/Weak patterns aids all-or-nothing matches on routine queries.

\item \textit{In-pipeline diagnostics confirm broad subtype mastery and degradation resilience.} On Uniform-300, models trained with \method{} supervision achieve near-ceiling accuracy on clean views and exhibit minimal degradation loss (\eg, only a 2.0-point drop for \textit{Real\,+\,Synthetic (\method{})} vs.\ 3.7 for \textit{Real\,+\,Replay} and up to 9.3 for zero-shot VLMs; Figure~\ref{fig:regime_diagnostics}(b)). While capture artifacts (blur, perspective warp) disrupt zero-shot recognition of small mathematical tokens, training on clean and degraded views generated under identical contracts encourages invariance to appearance shifts.
\end{itemize}

\subsection{Ablation and Supervision Quality Analysis}
\label{sec:analysis}

\paragraph{Construction ablation.}
To examine the contribution of each construction stage, we evaluate the three construction variants across 200 seed--subtype tasks (40 per subtype), spanning two layout/order subtypes (exact-same layout variant, option-order variant; 80 tasks) and three content subtypes (semantic-equivalent rephrase, extra-question containment, key-information difference; 120 tasks). We measure \emph{valid yield} (percentage of tasks producing a human-verified valid accepted pair) and \emph{false-admission rate} (FAR, percentage of accepted pairs judged invalid). This evaluation excludes the two-source hard-negative path.

\begin{figure}[htb]
    \centering
    \includegraphics[width=0.48\textwidth]{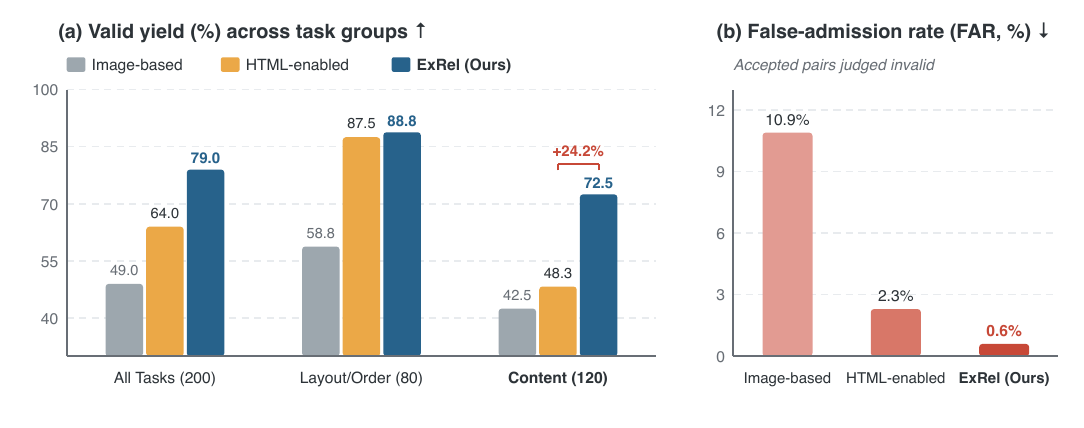}
    \caption{\textbf{Construction ablation study} across 200 tasks (Layout/Order: 80; Content: 120), comparing (a)~valid yield (\%) across task groups and (b)~false-admission rate (FAR, \%) among accepted pairs.}
    \label{fig:construction_ablation}
\end{figure}

Results are shown in Figure~\ref{fig:construction_ablation} (with full numerical breakdowns in Appendix~\ref{sec:app_qualitative}, Table~\ref{tab:construction_ablation_num}), from which we have the following observations:
\begin{itemize}[leftmargin=*,nosep]
\item \textit{Reconstructing an editable HTML/CSS substrate primarily resolves layout and ordering edits.} Moving from \emph{Image-based} to \emph{HTML-enabled} substantially boosts overall valid yield and sharply reduces FAR, with the improvement overwhelmingly concentrated on layout/order tasks while content yield remains low (Figure~\ref{fig:construction_ablation}(a)). Direct pixel editing frequently corrupts mathematical typesetting and glyph legibility, whereas exposing the DOM allows clean programmatic manipulation of CSS styles and option nodes, which suffices for layout rearrangements.

\item \textit{Executable relation contracts and verification are essential for reliable content interventions.} Integrating contract-guided execution and admission in \method{} further elevates overall valid yield and reduces false admissions to near zero (Figure~\ref{fig:construction_ablation}(b)), driven by a dramatic surge on content tasks (+24.2 points) while layout/order yield plateaus. Unconstrained HTML editing on content tasks frequently suffers from semantic drift---such as altering a core mathematical condition during rephrasing or inverting the containment direction. By locking invariant DOM slots ($\Phi_s$) and filtering out-of-contract edits via deterministic diffs and target-hidden review, \method{} prevents these semantic violations. Representative synthesized pairs across grades and subtypes are provided in Appendix~\ref{sec:app_qualitative} (Figure~\ref{fig:qualitative_examples}).
\end{itemize}

\paragraph{Audit of admitted supervision.}
We further examine label quality and acquisition cost in Table~\ref{tab:supervision_fidelity}. In a stratified human audit of 300 accepted \method{} pairs (20 per subtype), independent annotators confirm high post-admission validity at a substantially lower per-accepted-pair synthesis cost than outsourced annotation (0.70 vs.\ 1.10 RMB). By comparison, initial outsourced annotations on Release-200 show noticeably lower agreement with final adjudicated labels, reflecting the difficulty human annotators face when judging subtle numerical or directional differences without structural diffs. Because post-admission validity and pre-adjudication agreement follow different protocols, they characterize each source under its own workflow rather than a direct head-to-head comparison.

\begin{table}[htb]
\centering
\small
\caption{\textbf{Supervision quality and cost audit.}
Outsourced reports pre-adjudication agreement, whereas
\method{} reports post-admission validity under distinct
audit protocols.}
\setlength{\tabcolsep}{3pt}
\renewcommand{\arraystretch}{1.15}
\begin{tabular}{@{}lcccc@{}}
\toprule
\textbf{Source} & \textbf{Audited} &
\textbf{Protocol} & \textbf{Rate} &
\textbf{RMB/Pair} \\
\midrule
Outsourced & 1,000 &
Pre-adj.\ Agr. & 83.6 & 1.10 \\
\rowcolor{hlRow}
\textbf{\method{} (Ours)} & 300 &
Post-adm.\ Valid. & 98.0 & 0.70 \\
\bottomrule
\end{tabular}

\label{tab:supervision_fidelity}
\end{table}

\subsection{Downstream Utility}
\label{sec:downstream}

\paragraph{Grade-guided answer routing.}
We finally test whether predicted grades can route Release-200 queries during question answering. Under a fixed policy, a unique top-ranked Strong candidate triggers direct reuse of its archived answer; tied Strong candidates or a top grade of Medium trigger retrieval-augmented generation (RAG) with up to two top-tied candidates; and all-Weak predictions fall back to frozen query-only solving. We compare zero-shot Qwen3.5-4B (Base), \textit{Real\,+\,Replay}, and \textit{Real\,+\,Synthetic (\method{})} as routers across two untuned solvers, GPT-5.5~\cite{openai2026gpt55} and Qwen3.5-4B~\cite{qwen2026qwen35}.

\begin{table}[htb]
\centering
\caption{\textbf{Downstream answer routing (\%)} on Release-200. D / R / S: queries routed to direct reuse, RAG, and query-only solving.}
\footnotesize
\setlength{\tabcolsep}{3.6pt}
\renewcommand{\arraystretch}{1.15}
\begin{tabular}{lc|cc|cc}
\toprule
& & \multicolumn{2}{c|}{\textbf{GPT-5.5}} & \multicolumn{2}{c}{\textbf{Qwen3.5-4B}} \\
\cmidrule(lr){3-4}\cmidrule(lr){5-6}
\textbf{Router} & \textbf{D / R / S} & \textbf{Acc.} & $\boldsymbol{\Delta}$ & \textbf{Acc.} & $\boldsymbol{\Delta}$ \\
\midrule
Query-only & -- & 74.0 & -- & 74.0 & -- \\
Base & 60/129/11 & \underline{86.0} & \underline{+12.0} & \underline{86.5} & \underline{+12.5} \\
Real\,+\,Replay & 52/136/12 & \textbf{86.5} & \textbf{+12.5} & 85.0 & +11.0 \\
\rowcolor{hlRow}
\textbf{Real\,+\,Synthetic (\method{})} & 50/140/10 & 85.5 & +11.5 & \textbf{87.0} & \textbf{+13.0} \\
\bottomrule
\end{tabular}

\label{tab:downstream_routing}
\end{table}

As shown in Table~\ref{tab:downstream_routing}, all three grade-based routers substantially outperform query-only solving across both solvers, confirming that predicted grades provide effective routing signals for answer reuse and evidence selection. Compared with Base, the fine-tuned routers trigger direct reuse more conservatively and route more queries to RAG, reflecting stricter filtering of false-positive Strong matches. Meanwhile, the three routers perform within a narrow margin on both solvers, with the top router flipping across solvers. This narrow spread arises because the RAG branch acts as a tolerance buffer: whether a helpful candidate is graded as tied-Strong or Medium, capable solvers can still extract the relevant solution steps from the top-two retrieved candidates, so fine-grained boundary improvements are partially absorbed unless a router commits severe errors (such as promoting a Weak distractor to unique-Strong direct reuse).

\section{Conclusion}
\label{sec:conclusion}

We presented \method{}, a code-grounded agentic synthesis framework for learning directed reuse relevance between rich-text K--12 question images. By reconstructing real question images into editable HTML/CSS seeds and compiling 15 subtype definitions into executable relation contracts, \method{} couples precise structural and semantic interventions with deterministic and reviewer-assisted admission checks, while keeping downstream predictors strictly image-based. Across Release-200 and Uniform-300, \method{} supervision sharpens pair-level boundary discrimination---especially on scarce Medium relations---and improves degradation robustness over natural training and class-matched replay. Ablations, human audits, cost analyses, and answer-routing experiments further confirm that executable contracts raise valid synthesis yield, reduce false admissions and unit acquisition cost, and provide actionable routing signals for downstream question answering. Future work will extend contract-guided visual synthesis to broader document domains and multi-step pedagogical relations.


\bibliographystyle{ieeenat_fullname}
\bibliography{aaai2027}

\clearpage
\maketitlesupplementary
\appendix

\section{Contract-Guided Synthesis Algorithm}
\label{sec:app_taxonomy}

Algorithm~\ref{alg:exrel} details the three-stage synthesis workflow outlined in Section~\ref{sec:overview}: (1)~reconstructing a real question image $I^{\mathrm{real}}$ into an editable HTML/CSS seed $x=(I,H,m)$ via Image2HTML and profiling its structural metadata; (2)~routing eligible subtypes $\mathcal{E}(x;\mathcal{B})$ under precondition $P_s$ and instantiating the executable relation contract $\mathcal{C}_s$ into an execution plan $\pi$; and (3)~iteratively proposing, validating (via deterministic checks $V_s$ and target-hidden review $\tilde{y}$), and repairing candidate pairs within bounded budget $R_s$.

\begin{algorithm}[h]
\caption{\method{}: Contract-guided relation synthesis}
\label{alg:exrel}
\footnotesize
\renewcommand{\algorithmicrequire}{\textbf{Input:}}
\renewcommand{\algorithmicensure}{\textbf{Output:}}
\begin{algorithmic}[1]
\REQUIRE Image $I^{\mathrm{real}}$, target $\tau$, contracts $\{\mathcal{C}_s\}_{s\in\mathcal{S}}$, source pool $\mathcal{B}$
\ENSURE Accepted $\bigl((q,k,y_s),z\bigr)$ or rejection $(\bot,z)$

\STATE \textcolor{cvprblue}{\textit{\# Prepare executable seed}}
\STATE $H\gets\mathrm{Image2HTML}(I^{\mathrm{real}})$; \ $I\gets\mathrm{Render}(H)$
\STATE $m\gets\mathrm{Profile}(I,H)$; \ $x\gets(I,H,m)$

\STATE \textcolor{cvprblue}{\textit{\# Route subtype and instantiate contract}}
\STATE $\mathcal{E}\gets\{s\in\mathcal{S}\mid P_s(x;\mathcal{B})=1\}$
\STATE $s\gets\mathrm{SelectSubtype}(\mathcal{E},\tau)$
\IF{$s=\bot$}
  \STATE \textbf{return} $\bigl(\bot,\mathrm{RejectTrace}(x,\tau)\bigr)$
\ENDIF
\STATE $\pi\gets\mathrm{Instantiate}(\mathcal{C}_s,x)$; \ $z\gets\mathrm{InitTrace}(x,\pi)$

\STATE \textcolor{cvprblue}{\textit{\# Execute and verify directed relation}}
\WHILE{$\mathrm{WithinBudget}(z,R_s)$}
  \STATE $(q,k,e)\gets\mathrm{ProposePair}(x,\pi;\mathcal{B})$
  \STATE $\mathbf{v}\gets\mathrm{Check}_{V_s}(q,k,e)$
  \STATE $\tilde{y}\gets\mathrm{TargetHiddenReview}(q,k,e)$
  \STATE $a\gets\mathrm{Accept}_s(\mathbf{v},\tilde{y})$ \hfill $\triangleright$ Eq.~\eqref{eq:accept}
  \STATE $z\gets\mathrm{Log}(z,e,\mathbf{v},\tilde{y},a)$
  \IF{$a=1$}
    \STATE \textbf{return} $\bigl((q,k,y_s),z\bigr)$
  \ENDIF
  \STATE \textcolor{cvprblue}{\textit{\# Apply bounded feedback}}
  \STATE $\pi\gets\mathrm{ReviseOrStop}(\pi,z;R_s)$
  \IF{$\pi=\texttt{stop}$}
    \STATE \textbf{break}
  \ENDIF
\ENDWHILE
\STATE \textbf{return} $(\bot,z)$
\end{algorithmic}
\end{algorithm}

\section{Off-the-Shelf Ranking and End-to-End Relevance Modeling}
\label{sec:app_ranking}
\begin{figure*}[!t]
    \centering
    \includegraphics[width=0.96\textwidth]{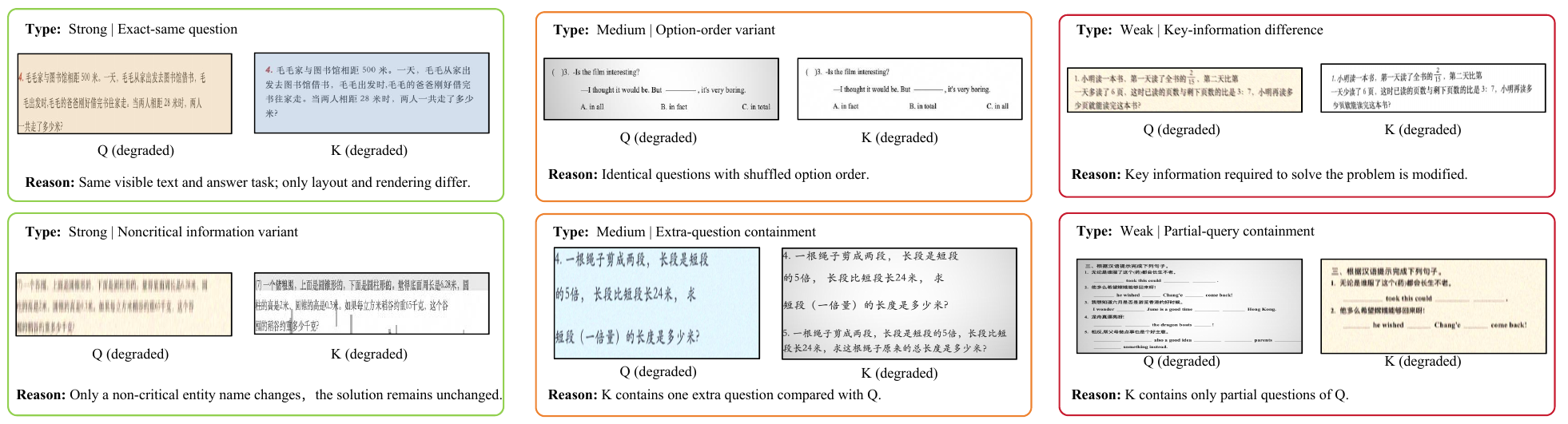}
    \caption{\textbf{Representative degraded pairs synthesized by \method{}} across Strong, Medium, and Weak subtypes, illustrating how contract-guided interventions and photographic degradation preserve task legibility, grade semantics, and query--candidate direction.}
    \label{fig:qualitative_examples}
\end{figure*}
\paragraph{Off-the-shelf ranking configurations and metrics.}
To complement the main evaluation in Section~\ref{sec:setup}, we evaluate off-the-shelf text-based and multimodal ranking models on Release-200. OCR-based ranking baselines operate on recognized question text and include Qwen3-Embedding-0.6B~\cite{zhang2025qwen3embedding}, Qwen3-Reranker-0.6B~\cite{zhang2025qwen3embedding}, BGE-M3~\cite{chen2024bge}, and BGE-Reranker-v2-m3~\cite{baai2024bgereranker}. Vision--language rerankers operate directly on ordered question-image pairs and include the 2B and 8B variants of Qwen3-VL-Reranker~\cite{li2026qwen3vlembedding}. We compare these ranking pipelines against zero-shot Qwen3.5-4B and Qwen3.5-27B~\cite{qwen2026qwen35}. In addition to three-grade tie-aware nDCG@5 (with gold gains $2,1,0$ for Strong, Medium, and Weak), we report Hit@1, MRR@5, and R@3 by treating Strong candidates as relevant.

\begin{table}[h]
\centering
\small
\setlength{\tabcolsep}{4.5pt}
\renewcommand{\arraystretch}{1.12}
\caption{\textbf{Off-the-shelf ranking performance (\%) on Release-200.} OCR-based models operate on recognized text, whereas vision--language models consume the original image pairs. Hit@1, MRR@5, and R@3 are computed for Strong candidates.}
\label{tab:off_the_shelf_ranking}
\begin{tabular}{@{}lcccc@{}}
\toprule
\textbf{Method} & \textbf{nDCG@5} & \textbf{Hit@1} & \textbf{MRR@5} & \textbf{R@3} \\
\midrule
\multicolumn{5}{@{}l}{\textit{OCR-based models}} \\
Qwen3-Embedding-0.6B & 90.29 & 72.58 & 82.50 & 78.36 \\
Qwen3-Reranker-0.6B & 76.53 & 35.65 & 59.30 & 62.92 \\
BGE-M3 & 88.49 & 66.94 & 80.12 & 80.44 \\
BGE-Reranker-v2-m3 & 88.06 & 67.74 & 79.78 & 75.20 \\
\midrule
\multicolumn{5}{@{}l}{\textit{Vision--language rerankers}} \\
Qwen3-VL-Reranker-2B & 89.50 & 66.13 & 80.39 & 81.18 \\
Qwen3-VL-Reranker-8B & 89.90 & 64.52 & 78.09 & 75.27 \\
\midrule
\multicolumn{5}{@{}l}{\textit{Zero-shot VLMs}} \\
Qwen3.5-4B & \underline{92.85} & \underline{78.70} & \underline{88.17} & \underline{88.04} \\
Qwen3.5-27B & \textbf{93.47} & \textbf{84.27} & \textbf{91.17} & \textbf{88.73} \\
\bottomrule
\end{tabular}
\end{table}

\paragraph{End-to-end relevance modeling vs.\ OCR/reranking pipelines.}
Table~\ref{tab:off_the_shelf_ranking} compares direct image-based relevance modeling with OCR-based embedding and reranking pipelines as well as dedicated vision--language rerankers. Both the 4B and 27B variants of Qwen3.5~\cite{qwen2026qwen35} consistently outperform all evaluated OCR-based pipelines and off-the-shelf vision--language rerankers across all four ranking metrics (\eg, reaching 92.85 and 93.47 nDCG@5, vs.\ 90.29 for the strongest OCR baseline). These results support modeling ordered question-image pairs end-to-end with a VLM, which preserves direct access to text, mathematical expressions, diagrams, and their spatial arrangement without suffering from cascading OCR recognition errors or layout loss. Building on this end-to-end formulation, \method{} provides targeted synthetic supervision that substantially improves fine-grained discrimination among Strong, Medium, and Weak relations (Table~\ref{tab:main_results}).

\section{Qualitative Examples and Construction Quality Breakdown}
\label{sec:app_qualitative}

\paragraph{Detailed numerical results of construction ablation.}
Table~\ref{tab:construction_ablation_num} provides the exact numerical breakdown corresponding to the construction ablation study in Section~\ref{sec:analysis} (Figure~\ref{fig:construction_ablation}) across the 200 seed--subtype tasks (80 Layout/Order and 120 Content tasks). While exposing the editable HTML/CSS substrate largely resolves layout and option-ordering interventions, \method{}'s executable contracts and evidence-aligned admission drive the primary gain on content-level interventions (+24.2 points in valid yield over unconstrained HTML editing) while reducing false admissions to near zero.

\begin{table}[h]
\centering
\small
\setlength{\tabcolsep}{4.5pt}
\renewcommand{\arraystretch}{1.12}
\caption{\textbf{Numerical breakdown of the construction ablation study (\%)} across 200 seed--subtype tasks (corresponding to Figure~\ref{fig:construction_ablation}). Valid yield is the percentage of tasks producing a human-verified valid accepted pair; FAR is the false-admission rate among accepted pairs.}
\label{tab:construction_ablation_num}
\begin{tabular}{@{}lcccc@{}}
\toprule
& \multicolumn{3}{c}{\textbf{Valid Yield (\%) $\uparrow$}} & \textbf{FAR (\%) $\downarrow$} \\
\cmidrule(lr){2-4}
\textbf{Configuration} & \textbf{All} & \textbf{Layout/Order} & \textbf{Content} & \textbf{Accepted} \\
\midrule
Image-based & 49.0 & 58.8 & 42.5 & 10.9 \\
HTML-enabled & 64.0 & 87.5 & 48.3 & 2.3 \\
\rowcolor{hlRow}
\textbf{\method{} (Ours)} & \textbf{79.0} & \textbf{88.8} & \textbf{72.5} & \textbf{0.6} \\
\bottomrule
\end{tabular}
\end{table}

\paragraph{Qualitative examples.}
Figure~\ref{fig:qualitative_examples} presents representative degraded question-image pairs synthesized by \method{} across six of the 15 relation subtypes spanning Strong, Medium, and Weak grades (referenced in Section~\ref{sec:analysis}). By executing subtype-specific relation contracts on reconstructed HTML/CSS seeds and applying controlled photographic degradation, \method{} preserves mathematical and diagram legibility, enforces target grade boundaries, and maintains the intended query--candidate direction $q\!\rightarrow\!k$.

\end{document}